\documentclass[11pt,fleqn]{article}
\usepackage[top=1in, bottom=1in, left=1in, right=1.5in]{geometry}

\usepackage{moreverb}
\usepackage{amsbsy}
\usepackage{amsmath}
\usepackage{authblk} 
\usepackage[document]{ragged2e} 
\usepackage{caption}
\usepackage{booktabs}
\usepackage{pdfpages}

\usepackage[colorlinks,bookmarksopen,bookmarksnumbered,citecolor=red,urlcolor=red]{hyperref}

\newcommand\BibTeX{{\rmfamily B\kern-.05em \textsc{i\kern-.025em b}\kern-.08em
T\kern-.1667em\lower.7ex\hbox{E}\kern-.125emX}}

\usepackage{enumitem}
\usepackage{xfrac}
\usepackage{csquotes}
\newcommand{\rowgroup}[1]{\hspace{-1em}#1}

\makeatletter
\renewcommand{\maketitle}{\bgroup\setlength{\parindent}{0pt}
\begin{flushleft}
  \textbf{\@title} \\[24pt]

  \@author
\end{flushleft}\egroup
}
\makeatother

\begin{document}


\title{\huge{Simulating realistically complex comparative effectiveness studies with time-varying covariates and right-censored outcomes}}

\author[1]{Maria E.~{Montez-Rath}}
\author[2]{Kristopher Kapphahn}
\author[2]{Maya B.~Mathur}
\author[2]{Natasha Purington}
\author[3]{Vilija R.~Joyce}
\author[2]{Manisha Desai\vspace{-2ex}}

\affil[1]{\small{\textit{Division of Nephrology, Stanford University School of Medicine, Palo Alto, CA, USA}}}
\affil[2]{\small{\textit{Quantitive Sciences Unit, Stanford University School of Medicine, Palo Alto, CA, USA}}}
\affil[3]{\small{\textit{VA Health Economics Resource Center, VA Palo Alto Health Care System, Menlo Park, CA, U.S.A.}}}

\renewcommand\Authands{ and }
{
  \renewcommand{\thefootnote}%
    {\fnsymbol{footnote}}
  \footnotetext[1]{Corresponding author: Maria E. Montez Rath, Division of Nephrology, Stanford University School of Medicine, 1070 Arastradero Road, Suite 3C11C, Palo Alto, CA 94304. Email: mmrath@stanford.edu}
}

\maketitle

\noindent
\textbf{Abstract.} Simulation studies are useful for evaluating and developing statistical methods for the analyses of complex problems. Performance of methods may be affected by multiple complexities present in real scenarios.  Generating sufficiently realistic data for this purpose, however, can be challenging. Our study of the comparative effectiveness of HIV protocols on the risk of cardiovascular disease -- involving the longitudinal assessment of HIV patients -- is such an example. The correlation structure across covariates and within subjects over time must be considered as well as right-censoring of the outcome of interest, time to myocardial infarction. A challenge in simulating the covariates is to incorporate a joint distribution for variables of mixed type -- continuous, binary or polytomous. An additional challenge is incorporating within-subject correlation where some variables may vary over time and others may remain static. To address these issues, we extend the work of Demirtas and Doganay (2012). Identifying a model from which to simulate the right-censored outcome as a function of these covariates builds on work developed by Sylvestre and Abrahamowicz (2007). In this paper, we describe a cohesive and user-friendly approach accompanied by R code to simulate comparative effectiveness studies with right-censored outcomes that are functions of time-varying covariates.  \par


\section{Introduction}
Simulation studies are often relied upon to demonstrate or evaluate properties of statistical methods when theoretical properties may be intractable.  When these properties are well characterized, they can be used with confidence. 

For example, while performing a comparative effectiveness study of antiretroviral therapies where the goal was to understand the comparative risk of cardiovascular disease among patients with HIV, co-authors of this team faced complex issues with missing data in a setting where many practical challenges had not yet been studied. One solution was to explore properties of estimators (bias and efficiency) across multiple methods for handling missing data under various conditions of missingness in a study that simulated a similarly complex longitudinal setting. In order to ensure confidence in the findings, the study should generate data that reflect those challenges encountered in the motivating data set including how missingness was induced.  In addition, other aspects of the underlying data should include: 1) variables that were correlated within a subject over time; 2) variables that were correlated across subjects; 3) variables of mixed type (continuous, binary, and categorical); 4) an outcome that was right-censored; and, 5) subjects with varying length of follow-up. 
 
In this paper, we present our methodology for simulating comparative effectiveness studies that, although developed with a specific motivating data set in mind, can be used to realistically simulate any comparative effectiveness study with a right-censored outcome.  We illustrate its use through one simulation study that closely mimics a real study, maintaining interrelationships among variables and across subjects. Our methods can be used in whole or in part, as needed, and we demonstrate their use in a power calculation for a comparative effectiveness study.

The paper is organized as follows. Section 2 presents our motivating data set, used in a comparative effectiveness study of antiretroviral therapies and cardiovascular disease. In Section 3, we provide background on the methods used in our approach, which is presented in Section 4. In Section 5, we evaluate the performance of our algorithm using commonly accepted accuracy and precision measures. In Section 6, we illustrate the utility of our approach for conducting a power calculation for a comparative effectiveness study. We provide a discussion and concluding remarks in Section 7. 

\section{Motivating Study}
Our work is motivated by a comparative effectiveness study of the risk of antiretroviral therapies (ART) on cardiovascular disease in US veterans with HIV. To address questions about comparative effectiveness, data from the Veterans Health Administration Clinical Case Registry (CCR) were used to create a longitudinal data set that compiled demographic, diagnostic, therapeutic, and health care utilization data on all HIV-infected patients from all VHA facilities \cite{desai2015}.  The final cohort included 24,510 HIV-infected patients from January 1996 through December 2009 exposed to at least one antiretroviral drug during that time period.  The outcome of interest was the occurrence of any cardiovascular event including MI, stroke, or cardiovascular procedures such as percutaneous coronary intervention and coronary artery bypass surgery. Subjects were followed from the date of the first positive HIV lab test or first ART prescription until the event of interest, death, or last documented activity in the CCR.  Subjects who did not develop the event of interest by the end of the study period (December 31, 2009) or who died during the study period without experiencing the event were censored. The exposures of interest were indicators of current use of each of 15 antiretroviral medications. These were coded as time-varying indicators, allowing for a new record to be created every time there was a medication switch. Time-invariant covariates included age, sex, and race (coded as white, black, and other). Time-varying covariates included body mass index (BMI) and viral load (VL). These were defined for those periods specified by changes in exposures to the drugs of interest. Because this data set -- created from multiple sources including pharmacy data, billing data, and clinical records -- posed unique challenges to the analysis while also representing how many comparative effectiveness studies are derived, we aimed to create a simulation tool that could maintain many of its complexities.

\section{Background}
In this section, we give an overview of the methods that provide the backbone of our approach for simulating studies that reflect our motivating data set. 

\subsection{Generation of Time-Varying Exposures and other Covariates}
We used methods found in the literature to generate the covariates and exposure variables. Specifically, we rely on the algorithm developed by Demirtas and Doganay to generate correlated binomial and normal data implemented in the R package \emph{binnor} \cite{demirtas2012,demirtas2014}. In this algorithm, the authors combined the framework of Emrich and Piedmonte to generate correlated binary data with an algorithm that generates multivariate normal data \cite{emrich1991,genz2010}. It requires the user to provide marginal characteristics such as means and variances as well as the association structure through correlations among all variables. These can be chosen arbitrarily or informed by a motivating data set. In essence, the mechanism assumes that all variables follow a multivariate normal density, but that some will be dichotomized. The algorithm then computes an overall correlation matrix that allows the generation of a multivariate normal distribution with pre-specified correlations. Binary variables are dichotomized by limits that are computed from the marginal proportions provided. See Demirtas and Doganay for more details \cite{demirtas2012}.

\begin{figure}
\centering
\resizebox*{18cm}{!}{\includegraphics{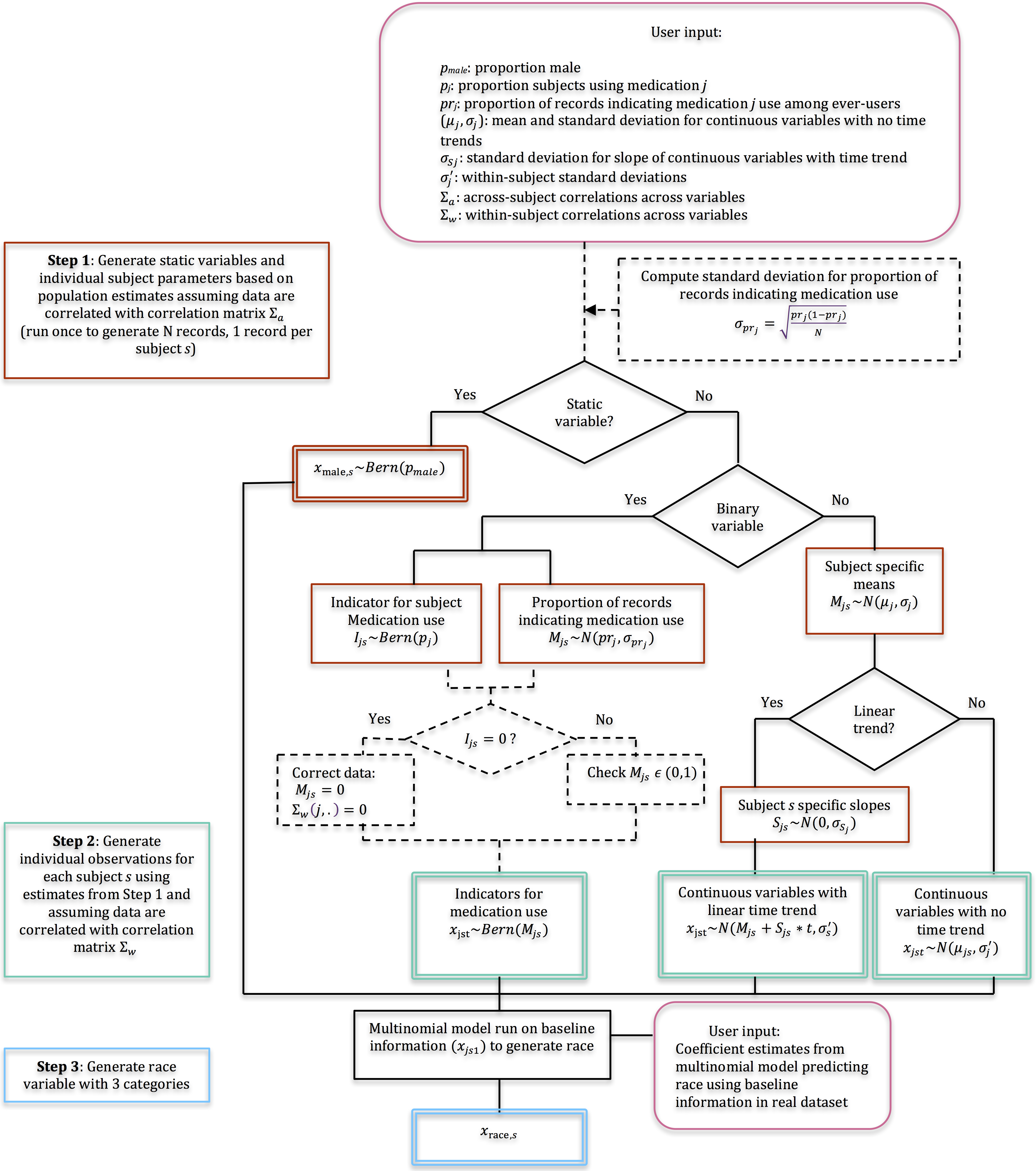}}
\caption{Description of the covariate generation algorithm. Boxes are color coded to show the output of each of the 3 steps. In step 1 (red), the algorithm outputs static variables and individual subject parameters; in step 2 (green), it outputs individual observations for each subject's time-varying variables; and, in step 3 (blue), it generates the polytomous variable. All user inputs are shown in pink boxes. Boxes with a dashed outline show internal checks.}
\label{fig1}
\end{figure}

We generated categorical data from a multinomial logistic regression and imposed a linear pattern over time within a subject on certain variables through the use of mixed effects models. Figure \ref{fig1} depicts the full algorithm; the various components will be described in more detail in Section 4.1.

\subsection{Generation of Outcome}
Several methods to generate survival times that follow a Cox model with time-varying covariates have been proposed in the literature (see Austin, Sylvestre, and Abrahamowicz) \cite{austin2012,sylvestre2008,abrahamowicz1996}. Of these, there are two methods that are relevant for our work in that they allow an arbitrary number of time-varying covariates with easy implementation. The first is a method developed by Zhou and implemented by Hendry \cite{zhou2001,hendry2014}. Zhou showed that right-censored outcomes can be generated from a transformation of piecewise exponential random variables, where the hazard is assumed constant within a time interval but can vary across time intervals that are defined by changes in the covariates. Although promising, we found that the current implementation of the algorithm created by Hendry can be biased in the presence of covariates that are correlated within a subject \cite{montezrath2017}. Hence, our data generating process relies on the second method -- proposed by Sylvestre and Abrahamowicz and implemented in the R package \emph{permalgo} \cite{sylvestre2008,sylvestre2015}. It is a permutation-based algorithm first introduced by Abrahamowicz \emph{et al.}, described in detail by MacKenzie and Abrahamowicz, and evaluated for use with time-varying covariates by Sylvestre and Abrahamowicz \cite{abrahamowicz1996, mackenzie2002, sylvestre2008}. 

\begin{figure}
    \centering
    \resizebox*{17cm}{!}{\includegraphics{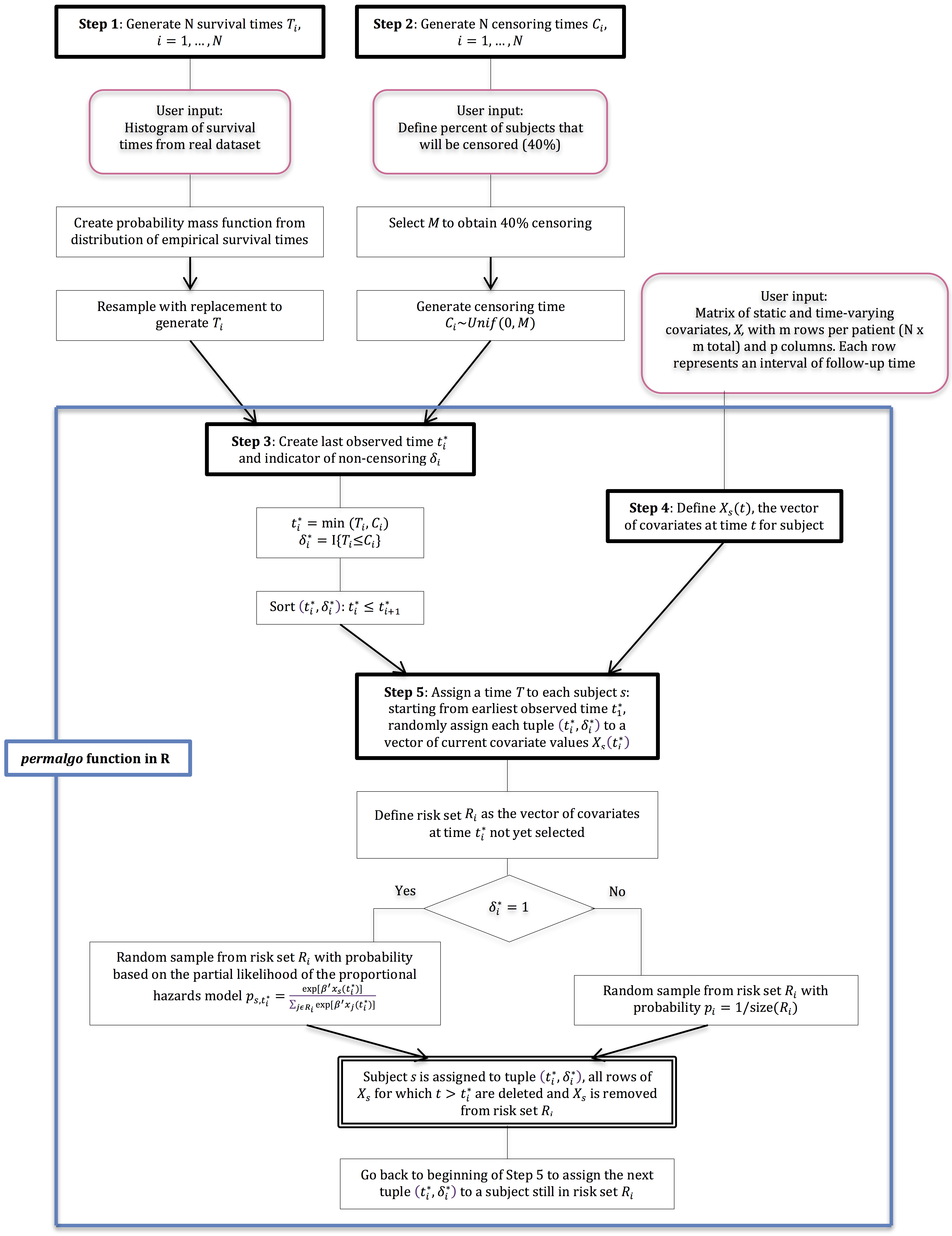}}
    \caption{Description of the algorithm to generate outcome: survival times that follow a Cox model with time-varying covariates. User inputs are shown in pink boxes.}
    \label{fig2}
\end{figure}

The Sylvestre and Abrahamowicz algorithm differs from many standard approaches that simulate structured random data from analytic expressions of a statistical model. Instead, this algorithm requires one to first specify survival times, independent covariates, and their relationship. From this, the independently generated times are matched to covariates, one-to-one, based on a permutation probability law derived from the partial likelihood of the Cox proportional hazards model \cite{abrahamowicz1996,mackenzie2002}. The algorithm, also depicted in Figure \ref{fig2}, consists of 5 main steps:

\begin{enumerate}[label*=\arabic*.]
    \item Generate $N$ survival times $T_i, i=1,\ldots,N$ from some user-specified distribution, assumed to represent the marginal distribution of survival times in the entire population;
    \item Independently from the survival times, generate N censoring times $C_i, i=1,\ldots,N$, also from some user-specified distribution;
    \item For each $i$, define the last observed time, $t_i^*=\mbox{min}(T_i,C_i)$ and an indicator of non-censoring, $\delta_i^*=I\{T_i<C_i\}$. Sort $(t_i^*,\delta_i^*)$ such that $t_i^* \leq t_{i+1}^*$;
    \item Provide (or generate) $N$ matrices with $p$ time-invariant and time-varying covariates and $m$ rows that represent intervals of follow-up time. Define $X_s(t)$ as the vector of covariates for subject $s$ at time $t$;
    \item Assign a time $T$ to each subject $s$. Start from earliest observed time $t_1^*$ and randomly assign each pair $(t_i^*,\delta_i^*)$ to a vector of current covariate values $X_s(t_i^*), s=1,\ldots,N$; Define the risk set for time $t_i^*$, $R_i$, as the vector of covariates at time $t_i^*$ for all the individuals not yet assigned a time $T$. The assignment depends on whether $t_i^*$ is a censoring time or not: 
    \begin{enumerate}[label*=\arabic*.]
    	\item If $t_i^*$ is a censoring time, $\delta_i^*=0$, then select an individual at random from $R_i$ with probability $p_i=\sfrac{1}{\mbox{size}(R_i)}$;
	\item If subject has the event, $\delta_i^*=1$, then sample individuals from  $R_i$  with probability
		\begin{equation}
			p_{s,t_i^*}=\frac{\mbox{exp}\lbrack\beta^{'} x_s(t_i^*)\rbrack}{\sum_{j \in R_i}\mbox{exp}\lbrack\beta^{'} x_j(t_i^*)\rbrack}
		\end{equation}
    \end{enumerate}
    \item[] The sampled individual is removed from further consideration, i.e. from $R_i$, and the algorithm moves to the next largest event time and repeats the procedure. The algorithm proceeds until all survival times have been matched to an individual.
\end{enumerate}

\section{Our Approach to Simulating Studies Using an External Data Source}
\subsection{Simulation of Covariates}
Our approach to simulating covariates is designed to capture several common features of comparative effectiveness data: 1) covariates should be correlated both across variables at a given time point and within each subject over time; 2) certain variables should change over time according to specified trajectories; and 3) covariates can include at least one multi-level categorical variable (e.g., race). Maintaining correlation structure in the presence of these complex covariates is an unresolved challenge that we addressed using the algorithm by Demirtas and Doganay \cite{demirtas2012}. Full algorithm descriptions for generation of covariates and generation of outcome are depicted in Figure \ref{fig1} and Figure \ref{fig2}, respectively.

Based on the motivating dataset, we aimed to realistically simulate correlated covariates, comprising eight continuous variables: age, BMI, CD4 count, systolic and diastolic blood pressure (BPS and BPD, respectively), HDL and LDL cholesterol and triglycerides; 16 binary variables (1 indicating sex and 15 indicating whether a subject was currently taking a given drug); and, 1 categorical variable: race (defined as black, white or other). All continuous covariates were simulated as normally distributed random variables, a simplifying assumption justified by empirical distributions in the motivating data set. For 4 of the 8 continuous variables (BMI, LDL, HDL, and CD4 count) we additionally allowed for linear time effects within a subject, modeled using random slopes generated for each subject. 

A central challenge of simulating repeated-measures data is to adequately capture both the across- and within-subject correlation and marginal structures. To address this problem, we used a two-level simulation process. In Step 1 (see Section 4.1.1), we used population parameters estimated across subjects to generate subject-specific random effects, resulting in a single row per subject. In Step 2 (see Section 4.1.2), we expanded each subject's data longitudinally by generating individual observations around these subject-level random effects. For example, in order to generate triglycerides, in Step 1, we sampled values from a Normal distribution with mean and standard deviation equal to the population mean and standard deviation of triglycerides found in the motivating data set to represent the subject-specific random effect. Then, for each subject, we generated repeated observations from another Normal distribution with mean equal to that of the values generated in Step 1.

Special considerations were needed to simulate drug status indicator variables, for which both 1) the overall proportion of ever-users of each drug and 2) the proportion of observations in which the user was currently on the drug (among ever-users) most closely resembled the motivating dataset. We explicitly parameterized both considerations by first simulating two variables for each of the 15 drugs: a binary variable corresponding to each subject's ever-use status (\textquote{exposed}) and a continuous variable corresponding to a subject's proportion of observations the drug had been prescribed (\textquote{with exposure}), conditional on a subject being exposed. These two variables were then used to simulate the final drug status indicators for each time point.  

Throughout, we use capital Greek letters to denote population parameters specified based on estimates from the motivating data set, capital Roman letters to denote subject-specific random effects generated in Step 1, and lowercase Roman letters to denote individual observations generated in Step 2. 

\subsubsection{Step 1: Simulation of Subject-Specific Random Effects}
In Step 1, we jointly simulated continuous and binary variables with joint correlation structure $\Sigma_a$, the across-subject correlation matrix derived from the correlations observed in the motivating data set (computed as Pearson correlations between all variables). The result was a dataset with a single row for each of $N$ subjects (Table \ref{table1}). 

\begin{table}[h]
	\caption{Example of data generated in Step 1: generate subject-specific means.}
	\centering
	\begin{tabular}{ccccccc}
	\toprule
	Subject ID	&	Male	&	Age	&	BMI	&	Triglycerides	&	Exposed to Abacavir	&	Proportion of records \\
		&		&	&	&		&	(yes/no)	&	indicating Abacavir use \\
		&		&	&	&		& (\textquote{exposed})	&	(\textquote{with exposure}) \\
	 \midrule
1	&	1	&	25	&	22	&	88	&	1	&	0.3	\\
2	&	0	&	50	&	30	&	120	&	0	&	0.1	\\
3	&	1	&	40	&	25	&	100	&	1	&	0.2	\\
	\bottomrule
	\end{tabular}
	 \label{table1}
\end{table}

\textbf{Normally Distributed Random Variables}. A total of 27 variables were simulated in this step as Normally distributed random variables: the 8 continuous variables of interest, 4 subject-specific slopes for the continuous variables with linear time effects, and the 15 variables representing the proportions of observations indicating drug exposure use (\textquote{with exposure}). To allow simultaneous simulation of binary and continuous variables (thus accurately capturing their intercorrelations), we assumed the latter followed a Normal distribution.

In general, we simulated all normally distributed random variables as:
\begin{equation}
M_{js}\sim N(\mu_j,\sigma_j)
\end{equation}
where $M_{js}$ is the mean of variable $j$ for subject $s$, $\mu_j$ is the population mean and $\sigma_j$ is the population standard deviation of variable $j$, the across-subject standard deviation. For the variables representing the proportion of observations (\textquote{with exposure}), we estimated the population standard deviation as:
\begin{equation}
\sigma_j=\sqrt{\frac{\mu_j(1-\mu_j)}{N}}
\end{equation}
We simulated slopes for the linear time-varying variables centered around 0 as:
\begin{equation}
S_{js} \sim N(0,\sigma_{S_j})
\end{equation}
where $\sigma_{S_j}$ is a pre-specified standard deviation.

\textbf{Binary Variables}. There were 16 binary variables in Step 1, including 15 ever-use indicators and 1 additional time-static binary variable of interest (e.g., sex) that was to remain static within a subject. We simulated all binary variables as:
\begin{equation}
I_{js} \sim Bern(p_j)
\end{equation}
where $I_{js}$ is the binary indictor for variable $j$ and subject $s$, and $p_j$ is the estimated population proportion (\textquote{with exposure}, or male) for variable $j$.

\textbf{Intermediary Housekeeping Step}. Before the variables generated in Step 1 can be used to generate within-individual observations in Step 2, several \textquote{housekeeping} procedures were necessary. At the end of Step 1, the data set contained \textquote{exposed} indicators as well as \textquote{with exposure} proportions for each drug. Before proceeding to Step 2, in which the \textquote{with exposure} proportions alone were used to generate within-individual observations, the \textquote{exposed} indicators were used to set the \textquote{with exposure} proportion to 0 for all subjects assigned to never-use (exposed=0). The \textquote{exposed} indicators were then discarded. Additionally, as the \textquote{with exposure} variables were generated from a Normal distribution, they could contain values outside the range of [0,1]. In these cases, out-of-bound values were set to the nearest bound (0 or 1).

In cases where the \textquote{with exposure} proportion had been set to 0 or 1, the corresponding non-diagonal entries of the within-subject correlation matrix ($\Sigma_w$) were overwritten to 0. Moreover, entries of $\Sigma_w$ may be out of bounds imposed by the maximum magnitude of correlation allowable between two variables when one is binary. In these cases, the entry was overwritten to the closest bound, defined by subject-specific parameters for the density of the binary variable (Table \ref{table2}). 

\begin{table}[h]
	\caption{Example of data generated in Step 1 after \textquote{housekeeping} procedures.}
	\centering
	\begin{tabular}{ccccccc}
	\toprule
	Subject ID	&	Male	&	Age	&	BMI	&	Triglycerides	&	Exposed to Abacavir	&	Proportion of records \\
		&		&	&	&		&	(yes/no)	&	indicating Abacavir use \\
		&		&	&	&		& (\textquote{exposed})	&	(\textquote{with exposure}) \\
 \midrule
1	&	1	&	25	&	22	&	88	&	1	&	0.3	\\
2	&	0	&	50	&	30	&	120	&	0	&	0	\\
3	&	1	&	40	&	25	&	100	&	1	&	0.2	\\
	\bottomrule
	\end{tabular}
	 \label{table2}
\end{table}

\subsubsection{Step 2: Simulation of Individual Observations within Each Subject}
In Step 2, we used the subject-specific random effects generated in Step 1 jointly with the correlation structure $\Sigma_w$ to simulate data for each subject. The use of a new correlation matrix (distinct from matrix $\Sigma_a$ from Step 1) allows the correlation structure of individual observations within a subject to be controlled independently from the correlation structure across variables and subjects. This simulation process occurred once for each subject. 

\textbf{Normally Distributed Random Variables}. Within each subject, we generated individual observations for the 4 continuous variables of interest that do not vary as a function of time as:
\begin{equation}
x_{jst} \sim N(M_{js},\sigma_{js}^{'})
\end{equation}
where $x_{jst}$ represents the value of variable $j$ for subject $s$ at time $t$. $M_{js}$ is the mean obtained in Step 1 and $\sigma_{js}^{'}$ is the within-subject standard deviation of variable $j$ (fixed across all subjects and equal to 1/3 of the population standard deviation, $\sigma_j^{'}$). 

We generated the 5 continuous variables that vary as a function of time as:
\begin{equation}
x_{jst} \sim N(M_{js}+S_{js}*t,\sigma_j^{'})
\end{equation}
 where $\sigma_j^{'}$ is the error standard deviation for variable $j$. 

\textbf{Binary Variables}. We simulated 15 time-varying indicators for current exposure to drug as:
\begin{equation}
x_{jst} \sim Bern(M_{js})
\end{equation}
The remaining binary variable, sex, was static. Therefore the indicator for sex -- generated from Step1 -- was expanded to be a vector constant over the number of observations within a subject. Table \ref{table3} shows an example of what the data may look like at the end of this step.

\begin{table}[h]
	\caption{Example of covariate data generated in Step 2.}
	\centering
	\begin{tabular}{ccccccc}
	\toprule
Subject ID	&	Time	&	Male	&	Age	&	BMI	&	Triglycerides	&	Current exposure to 	\\
& & & & & & Abacavir (yes/no) \\
 \midrule
1	&	1	&	1	&	25	&	21.8	&	85	&	1	\\
1	&	2	&	1	&	25	&	21.7	&	95	&	0	\\
\dots	&	&		&		&		&		&		\\
1	&	200	&	1	&	25	&	22.3	&	87	&	1	\\  
& & & & & & \\
2	&	1	&	0	&	50	&	30.4	&	110	&	0	\\
2	&	2	&	0	&	50	&	32.1	&	118	&	0	\\
\dots	&	&		&		&		&		&		\\
2	&	200	&	0	&	50	&	29.6	&	130	&	0	\\
\dots	&	&		&		&		&		&		\\
	\bottomrule
	\end{tabular}
	 \label{table3}
\end{table}

\subsubsection{Step 3: Simulation of Categorical Variables}
Simulating a categorical variable such as race with 3 categories posed a special challenge. Simply including dummy variables to represent a multi-level categorical variable in the simulation step with the other binary and continuous covariates does not preserve the inherent collinearity of these variables. An obvious option is to jointly simulate all but one reference level of the indicator variable for race, but such a strategy may limit control over marginal probabilities and it would not preclude a subject's having more than one race. Therefore, rather than simulating the categorical variables representing race jointly with the other covariates, we simulated it via a multinomial model as a function of \textquote{baseline} covariates or those covariates that describe the subject at the first time of observation. We used \textquote{baseline} covariates because our approach was to include a mutually exclusive categorical variable that was static.  However, one could easily extend this to simulate a mutually exclusive categorical variable that is time-varying by deriving the term as a function of time-varying covariates. The parameters for these models were informed by the motivating data set.  

\subsection{Simulation of Time-Dependent Outcome}
For realism, we specified the distribution of the simulated survival times to closely resemble that of the observed. The algorithm uses a unit-less time measurement allowing for flexibility in the number of time points generated. This number should be sufficiently large to allow for granularity while not posing computational limitations. In the motivating data set, we observed a maximum time of 5112 days. However, for practical reasons we chose to only simulate data with 200 time points. To ensure comparable distributions of survival times in our motivating and simulated data sets, we rescaled the event times in our motivating data set by dividing each observed time by 200 (5112 / 200 = 25.5). After scaling, each event time was rounded up to the nearest integer, resulting in an appropriately scaled version of the event times from our motivating data set. The distribution of these scaled event times was used to construct a probability mass function (PMF) that corresponds to the desired event times. The PMF is a non-parametric distribution of the survival times in the original population. For each simulated data set, we sampled with replacement from this PMF to obtain N=2000 survival times. Censoring times were generated from a uniform distribution with parameters chosen so that 40\% of observations were censored. These survival and censoring times were used in Steps 1 and 2 of the algorithm used for outcome generation, respectively, described in Section 3.2 and depicted in Figure \ref{fig2}.

\section{Results}
We generated 2000 replications with 200 observations for each of 2000 subjects. In Tables 4 and 5, we present how closely our simulations match the motivating data set with regard to distributional properties by showing the empirical values (from the motivating data set), the average estimates across the 2000 replications (simulated), and the difference between the empirical and simulated quantities, for exposure variables (Table \ref{table4}) and covariates (Table \ref{table5}). For the 15 binary variables representing drug exposure (proportion exposed), the distance from the empirical values to the simulated quantities ranged from 0 (for 5 out of the 15 drugs) to 0.016. Differences between the empirical and average estimates for the proportion of records with exposure ranged in absolute value from 0.036 for the drug with lowest proportion (Saquinavir; true=0.04) to 0.13 for the drug with the largest proportion (Lamivudine; true=0.4) (Table \ref{table4}). Accuracy of covariates depended on whether the variable was generated as time-varying or fixed at baseline. We found that the algorithm generated fixed covariates with minimal differences between empirical and average estimated values (0 for the proportion of male sex, 0.109 for white race, 0.114 for black race and 0.003 for race other, and 0.008 for mean age and -0.012 for age standard deviation) while results for time-varying covariates were more variable. The algorithm generated data with means that were close to the empirical means (absolute distance from empirical value ranged from 0 for triglycerides to 0.208 for CD4 count). However, the generated time-varying data had standard deviations that were farther from the empirical standard deviations (on average) (Table \ref{table5}). 

\begin{table}[!htbp]
	\caption{Accuracy results for the exposures: proportion of patients who are exposed to a drug (\textquote{Proportion exposed}) and proportion of records where a drug is prescribed (\textquote{Proportion of records with exposure}).}
	\centering
	\begin{tabular}{lrrr}
	\toprule
Drug	&	Empirical 	&	Average	&	Distance from \\
	&	 Value	&	estimate &	 empirical	value \\
\midrule
\rowgroup{\textbf{Proportion exposed}}			\\	
Abacavir	&	0.275	&	0.291	&	0.016	\\
Atazanir	&	0.222	&	0.223	&	0.001	\\
Didanosine	&	0.198	&	0.198	&	0	\\
Efavirenz	&	0.500	&	0.500	&	0	\\
Emtricitabine	&	0.413	&	0.413	&	0	\\
Indinavir	&	0.206	&	0.214	&	0.008	\\
Lamivudine	&	0.688	&	0.673	&	-0.015	\\
Lopinavir	&	0.253	&	0.254	&	0.001	\\
Nelfinavir	&	0.240	&	0.244	&	0.004	\\
Nevirapine	&	0.184	&	0.185	&	0.001	\\
Ritonavir	&	0.180	&	0.180	&	0	\\
Saquinavir	&	0.103	&	0.105	&	0.002	\\
Stavudine	&	0.332	&	0.334	&	0.002	\\
Tenofovir	&	0.526	&	0.526	&	0	\\
Zidovudine	&	0.544	&	0.543	&	-0.001	\\
\rowgroup{\textbf{Proportion of records with exposure}}			\\
Abacavir	&	0.12	&	0.033	&	-0.087	\\
Atazanir	&	0.08	&	0.018	&	-0.062	\\
Didanosine	&	0.09	&	0.018	&	-0.072	\\
Efavirenz	&	0.16	&	0.080	&	-0.080 	\\
Emtricitabine	&	0.11	&	0.046	&	-0.064	\\
Indinavir	&	0.07	&	0.015	&	-0.055	\\
Lamivudine	&	0.40	&	0.270	&	-0.13	0 \\
Lopinavir	&	0.12	&	0.031	&	-0.089	\\
Nelfinavir	&	0.08	&	0.020	&	-0.060 	\\
Nevirapine	&	0.07	&	0.013	&	-0.057	\\
Ritonavir	&	0.05	&	0.009	&	-0.041	\\
Saquinavir	&	0.04	&	0.004	&	-0.036	\\
Stavudine	&	0.17	&	0.058	&	-0.112	\\
Tenofovir	&	0.22	&	0.116	&	-0.104	\\
Zidovudine	&	0.24	&	0.131	&	-0.109	\\
	\bottomrule
	\end{tabular}
	 \label{table4}
\end{table}

Plots of simulated trajectories for two time-varying variables (changing over time within a patient; A: BMI, B: LDL) for selected patients are shown in Supplemental Figure S1. These plots illustrate that the algorithm generated time-varying covariates with realistic trajectories with different levels of variability over time depending on the characteristics imposed and as expected. Plots depicting the correlation among variables in the generated data (Supplemental Figure S2, right) and correlations present in the motivating data set (Supplemental Figure S2, left) are shown in Supplemental Figure S2 and showed a comparable correlation structure, as desired.

\begin{table}[h]
	\caption{Accuracy results for fixed/time-varying covariates.}
	\centering
	\begin{tabular}{llrrr}
	\toprule
Covariate	&		&	Empirical 	&	Average 	&	Distance from 	\\
	&		&	value	&	estimate	&	empirical value	\\ 
\midrule
Male	&	\emph{p}	&	0.97	&	0.97	&	0	\\
	&		&		&		&		\\
Race	&		&		&		&		\\
  white	&	\emph{p}	&	0.338	&	0.447	&	0.109	\\
  black	&	\emph{p}	&	0.424	&	0.538	&	0.114	\\
  other	&	\emph{p}	&	0.012	&	0.015	&	0.003	\\
	&		&		&		&		\\
Age	&	$\mu$	&	46	&	46.008	&	0.008	\\
	&	$\sigma$	&	10.12	&	10.108	&	-0.012	\\
	&		&		&		&		\\
BMI	&	$\mu$	&	25	&	25.003	&	0.003	\\
	&	$\sigma$	&	4.2	&	4.974	&	0.774	\\
	&		&		&		&		\\
CD4 Count	&	$\mu$	&	308.09	&	307.882	&	-0.208	\\
	&	$\sigma$	&	241.54	&	278.859	&	37.319	\\
	&		&		&		&		\\
Systolic BP	&	$\mu$	&	126.85	&	126.866	&	0.016	\\
	&	$\sigma$	&	15.48	&	17.876	&	2.396	\\
	&		&		&		&		\\
Diastolic BP	&	$\mu$	&	76.7	&	76.703	&	0.003	\\
	&	$\sigma$	&	9.96	&	11.502	&	1.542	\\
	&		&		&		&	0	\\
LDL	&	$\mu$	&	99.1	&	99.104	&	0.004	\\
	&	$\sigma$	&	23	&	27.206	&	4.206	\\
	&		&		&		&		\\
HDL	&	$\mu$	&	39.02	&	39.012	&	-0.008	\\
	&	$\sigma$	&	7	&	9.004	&	2.004	\\
	&		&		&		&		\\
Triglycerides	&	$\mu$	&	180.87	&	180.87	&	0	\\
	&	$\sigma$	&	30	&	34.632	&	4.632	\\	
\bottomrule
\multicolumn{5}{l}{$p$, proportion; $\mu$, mean; $\sigma$, standard deviation; BP, blood pressure} \\
	\end{tabular}
	 \label{table5}
\end{table}

\begin{figure}
\centering
\resizebox*{18cm}{!}{\includegraphics{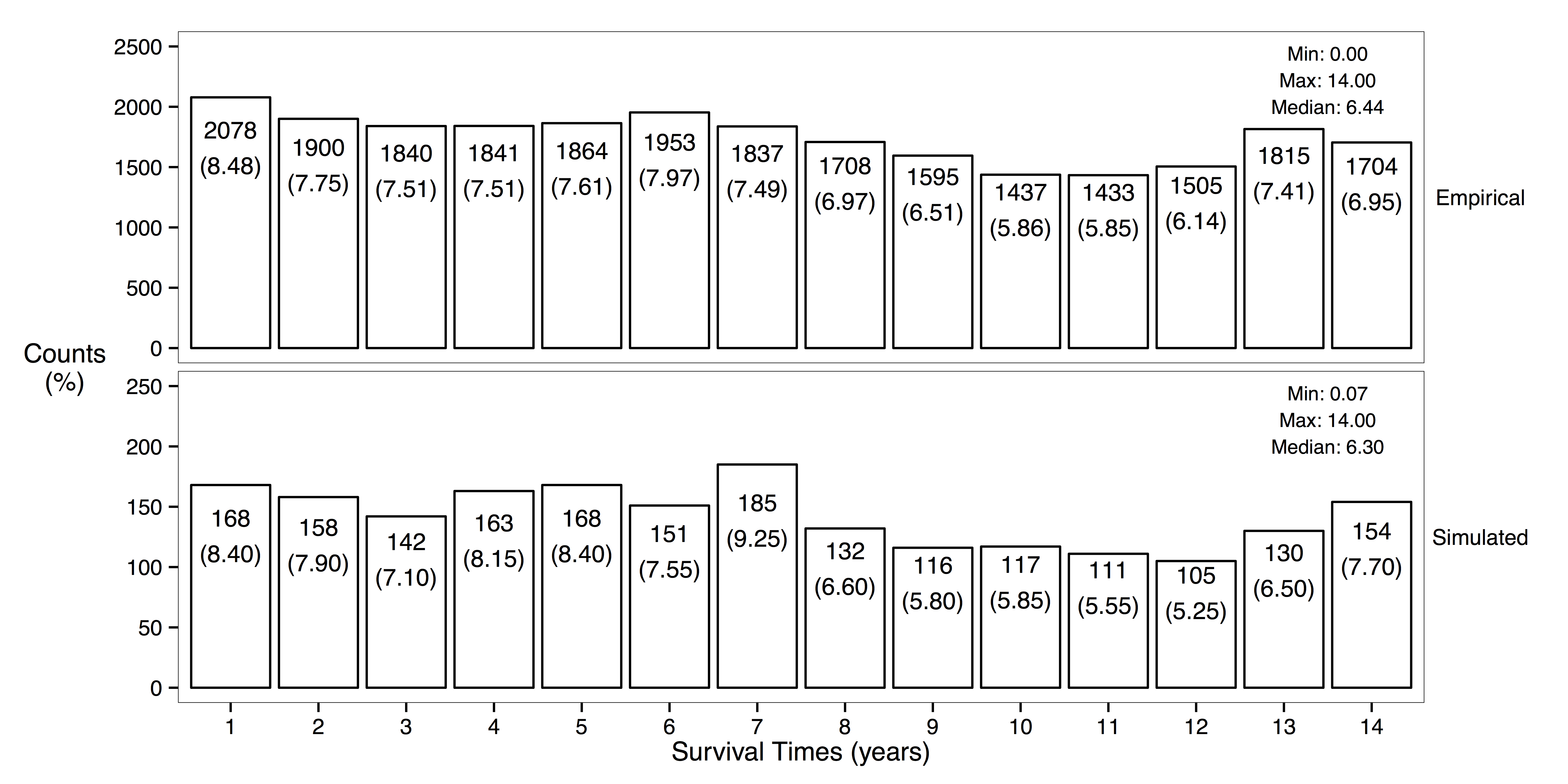}}
\caption{Distribution of time to event in the real data (empirical) and simulated data. p-value for Chi-square test comparing the two distributions equaled to 0.3.}
\label{fig5}
\end{figure}

The distributions of the right-censored outcome between the empirical and simulated data sets were also comparable. Figure \ref{fig5} depicts the two distributions for one simulated data set (p-value for Chi-square test comparing the two distributions = 0.3). Accuracy and precision measures based on the parameters used in the generation of survival times are shown in Table \ref{table6}. Average estimates across the 2000 replications were close to the true or targeted values, leading to relative and standardized biases that are negligible. RMSE values appear to be very small and coverage probabilities are all close to the nominal level of 95\%. 

\begin{table}[!htbp]
	\caption{Accuracy and precision measures in the generation of right-censored outcome.}
	\centering
	\begin{tabular}{lrrrrrrr}
	\toprule
	&	True	&	\multicolumn{6}{c}{Estimated log-hazard ratio} 	\\ \cline{3-8}	
	&	log-hazard ratio 	&	Mean	&	Bias	&	Avg. SE	&	Std. Bias	&	MSE	&	Coverage	\\
	\midrule
\rowgroup{\textbf{Indicator of drug exposure}} \\
Abacavir	&	0.7	&	0.686	&	-0.014	&	0.119	&	-12	&	0.0145	&	95\%	\\
Atazanir	&	0.5	&	0.484	&	-0.016	&	0.175	&	-9	&	0.0326	&	95\%	\\
Didanosine	&	0.3	&	0.285	&	-0.015	&	0.194	&	-8	&	0.0344	&	96\%	\\
Efavirenz	&	0	&	-0.005	&	-0.005	&	0.108	&	-5	&	0.0116	&	95\%	\\
Emtricitabine	&	0	&	-0.009	&	-0.009	&	0.148	&	-6	&	0.0229	&	95\%	\\
Indinavir	&	0	&	-0.026	&	-0.026	&	0.228	&	-11	&	0.0556	&	94\%	\\
Lamivudine	&	0.5	&	0.495	&	-0.005	&	0.070	&	-8	&	0.0050	&	95\%	\\
Lopinavir	&	0	&	-0.015	&	-0.015	&	0.166	&	-9	&	0.0284	&	95\%	\\
Nelfinavir	&	0	&	-0.011	&	-0.011	&	0.194	&	-6	&	0.0379	&	96\%	\\
Nevirapine	&	0	&	-0.029	&	-0.029	&	0.237	&	-12	&	0.0615	&	95\%	\\
Ritonavir	&	0	&	-0.040 &	-0.040	 &	0.301	&	-13	&	0.0969	&	96\%	\\
Saquinavir	&	0	&	-0.079	&	-0.079	&	0.438	&	-18	&	0.2199	&	96\%	\\
Stavudine	&	0	&	-0.005	&	-0.005	&	0.122	&	-4	&	0.0143	&	95\%	\\
Tenofovir	&	0	&	-0.005	&	-0.005	&	0.097	&	-5	&	0.0088	&	96\%	\\
Zidovudine	&	0	&	-0.002	&	-0.002	&	0.088	&	-2	&	0.0081	&	95\%	\\
\rowgroup{\textbf{Covariates}}			\\
\rowgroup{Male	} &	0.15	&	0.138	&	-0.012	&	0.192	&	-6	&	0.0381	&	95\%	\\
\rowgroup{Race} \\
White	&	Reference	&		&		&		&		&		&		\\
Black	&	-0.2	&	-0.202	&	-0.002	&	0.060	&	-4	&	0.0034	&	96\%	\\
Other	&	0	&	-0.026	&	-0.026	&	0.247	&	-11	&	0.0693	&	95\%	\\
Age	&	0.02	&	0.02	&	0	&	0.003	&	-13	&	0	&	95\%	\\
\rowgroup{BMI}		\\
$<20$	&	2	&	1.971	&	-0.029	&	0.088	&	-33	&	0.0096	&	91\%	\\
$20-25$	&	Reference	&		&		&		&		&		&		\\
$25-30$	&	0	&	-0.002	&	-0.002	&	0.089	&	-2	&	0.0081	&	94\%	\\
$>30$	&	2	&	1.972	&	-0.028	&	0.089	&	-32	&	0.0096	&	93\%	\\
\rowgroup{CD4 Count}		\\
$<50$	&	Reference	&		&		&		&		&		&		\\
$50-100$	&	0.15	&	0.148	&	-0.002	&	0.156	&	-1	&	0.0261	&	94\%	\\
$100-200$	&	-0.06	&	-0.057	&	0.003	&	0.119	&	2	&	0.0159	&	94\%	\\
$200-350$	&	0.03	&	0.026	&	-0.004	&	0.103	&	-4	&	0.0112	&	94\%	\\
$350-500$	&	-0.06	&	-0.065	&	-0.005	&	0.091	&	-5	&	0.0092	&	94\%	\\
\rowgroup{Blood Pressure} \\
Systolic 	&	0.005	&	0.005	&	0	&	0.003	&	-4	&	0	&	95\%	\\
Diastolic	&	0.005	&	0.005	&	0	&	0.002	&	-3	&	0	&	95\%	\\
\rowgroup{Cholesterol} \\
LDL	&	0.005	&	0.005	&	0	&	0.001	&	-12	&	0	&	93\%	\\
HDL	&	0.002	&	0.002	&	0	&	0.004	&	-2	&	0	&	94\%	\\	
	\bottomrule
	\end{tabular}
	 \label{table6}
\end{table}

\section{Application -- Simulation Tool for Designing Comparative Effectiveness Studies }
Off-the-shelf power calculators aid in designing studies by estimating the power (or sample size needed) for a single hypothesis test through closed-form equations that typically make an assumption about the relationship between one variable of interest and outcome. Our simulation tool, however, provides the flexibility to estimate other quantities, such as the probability of detecting a group of effects, while simultaneously allowing the inclusion of a more realistically complex set of features. 

We provide the following illustration of the utility of our method for a power calculation addressing a complex set of questions. Suppose investigators are interested in doing a comparative effectiveness study of HIV medications and have access to published data of a source such as the CCR. This information can be used to determine the power to identify which drug exposures of the 15 drugs considered are associated with cardiovascular events in the presence of multiple time-varying confounders including the drug exposures.  To illustrate, we assumed that some drugs (5) were associated with cardiovascular events with varying magnitudes of association, and that some were not.  Suppose the range of association was from modest to high magnitudes with log-hazards ranging from 0.05 to 0.7 (or hazard ratios ranging from 1.05 to 2.0), and with varying frequencies of exposure, where the latter was informed by the empirical data set.  We further assumed that the number of cardiovascular events was unknown but anticipated to be relatively low according to external data (occurring at a rate of 5\%).  A simulation was performed that generated 100 data sets with 20,000 subjects each and 150 observations per subject, with the following characteristics:
\begin{enumerate}[label*=\arabic*.]
	\item Exposures: 15 drugs  
	\begin{enumerate}[label*=\arabic*.]
		\item 50\% of patients were exposed to each of the 15 drugs
		\item 5 drugs were positively associated with outcome where prevalences of exposure are: 0.05, 0.1, 0.2, 0.3, and 0.4
		\item The remaining10 drugs were not associated with outcome and were generated with a prevalence of 0.2
	\end{enumerate}
	\item Covariates: All other covariates were generated using distributional properties informed by the motivating data set 
	\item Outcome: time to cardiovascular disease
	\begin{enumerate}[label*=\arabic*.]
		\item Time to cardiovascular disease was drawn from a Weibull distribution with parameters $\kappa=1.5$ (shape) and $\lambda=76.6$ (scale) for a median time-to-cardiovascular-event of 60 months
		\item Censoring times were drawn from a Weibull distribution, with parameters $\kappa=1$ (shape) and $\lambda=10$ (scale) to produce 95\% censoring
		\item Associations between other confounders and outcome were assumed to have characteristics as those described in Table \ref{table6}
		\item The 5 main drugs were assigned an effect that corresponded to hazard ratios (HRs) of 1.05, 1.15, 1.20, 1.50, and 2.00. The corresponding coefficients were permuted with drug prevalence resulting in 120 (5!) arrangements of the effects across the 5 drugs.
		\item The remaining 10 drugs were assumed to be unassociated with outcome with corresponding HRs of 1.0.
	\end{enumerate}
\end{enumerate}
To illustrate, for each possible permutation of effects, we generated 100 data sets and fit a Cox frailty model to each, assuming a varying baseline hazard for each patient, thus taking into account that each patient has multiple observations, due to changes in medication regimen. Using the simulated data sets, we computed the probability of detecting associations between outcome and all 5 drugs, the probability of detecting at least one association, as well as the probability of falsely detecting an association. Tests were conducted at the .05/15=0.003 level to account for multiplicity.

For comparison, we also computed the power using two standard software packages available in Stata: \emph{stpower cox} and \emph{stpower exponential}. \emph{stpower cox} assumes a Cox proportional hazards model and allows one to incorporate the expected correlation between the drug being tested and other covariates that will be included in the Cox PH model. \emph{stpower exponential} assumes that the survival times are exponentially distributed and that the observations are independent and no confounding is present.  For the former, we assumed a correlation between the drug of interest and other covariates of 0.2 on a data set with a sample size of 20,000. For both calculations, we assumed that each patient had one single record that contained both an indicator for drug use and outcome. Similar to the simulation, we conducted separate hypothesis tests on each of the 15 drug coefficients, each at the 0.003 level of significance. 

\subsection{Results from the Power Calculation Using the Simulation Tool}
\begin{figure}
\centering
\resizebox*{16cm}{!}{\includegraphics{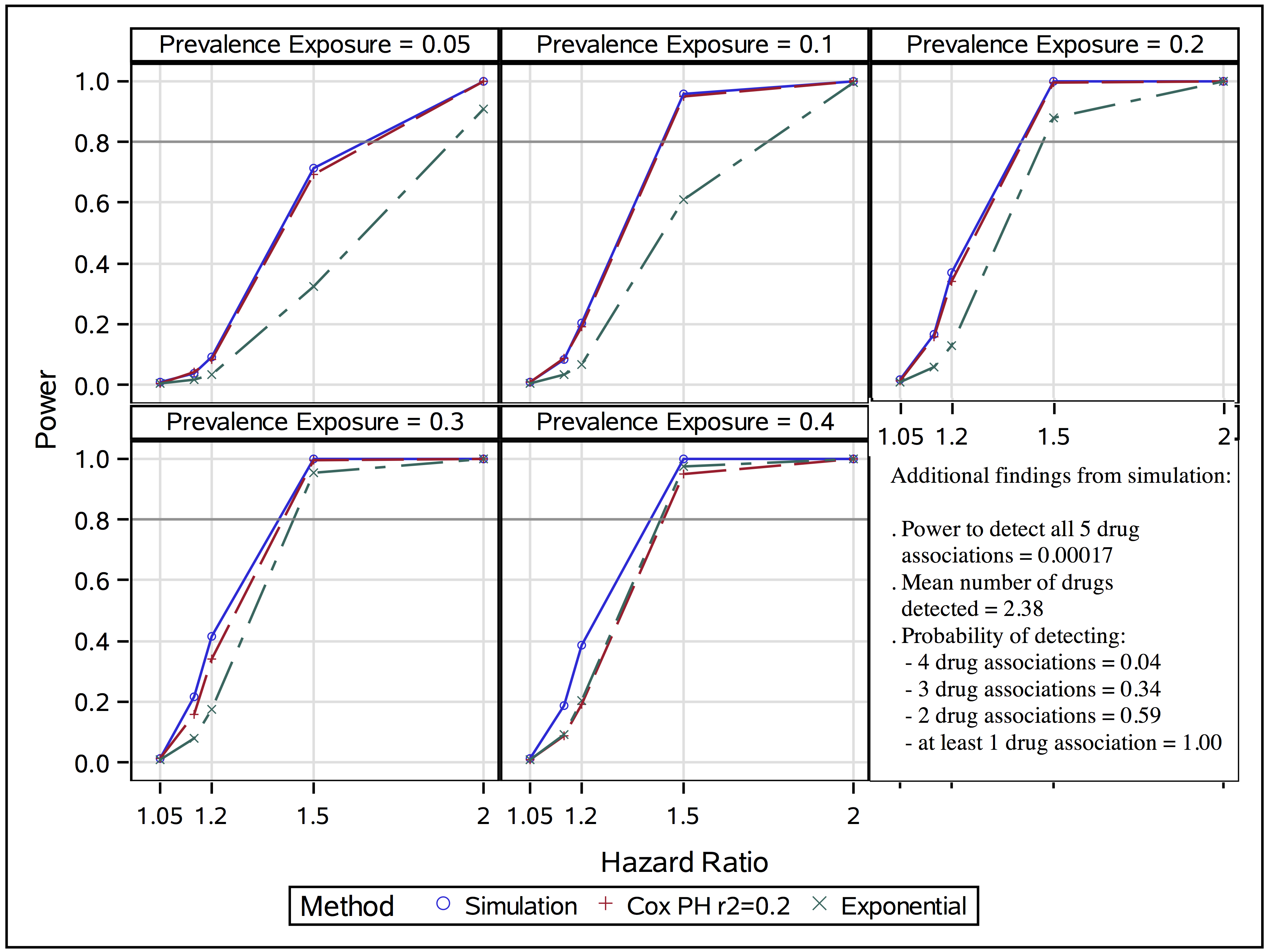}}
\caption{Comparison of power calculation using the simulation tool and Stata commands \emph{stpower cox} (Cox PH r2=0.2; assuming correlation between covariates to be equal to 0.2) and \emph{stpower exponential} for varying prevalences and hazard ratios}
\label{fig6}
\end{figure}

The simulation shows that with 20,000 subjects, we had little power to detect all associations of interest (0.00017). Even though the mean number of drugs detected across all data sets was 2.38, the probability of detecting associations with 2 drugs was only 0.6. As expected, the power decreased for drugs with lower prevalences. For a drug with a prevalence of 20\%, there was less than 40\% power to detect a hazard ratio of 1.2 but 100\% power to detect a hazard ratio of 1.5. For the lowest prevalence assumed (5\%), the power to detect a hazard ratio of 1.5 was 71\%. We observed similar results for the power to detect each drug individually using the \emph{stpower cox} command, however, the \emph{stpower exponential} command gave discrepant results, reporting lower power overall. Agreement in results across all 3 methods increased as exposure prevalence increased (Figure \ref{fig6}). Importantly, our simulation tool allowed evaluation of joint discoveries (Figure 4, bottom right panel) whereas the traditional tools do not yield such results.

\section{Discussion}
Simulation studies provide a means to evaluate methods, particularly when theoretical descriptions of statistical properties of estimators are not feasible \cite{peduzzi1996,demirtas2003,dormann2013,wan2015,royston2014}. For example, statistical properties of estimators may be well characterized only when the sample size reaches asymptotic levels, but have unknown behavior for smaller sample sizes \cite{roussos1996,vanbuuren2007}.  Such studies are increasing in importance in the current era of big data, where complexities such as an unknown correlation structure of a large number of features may affect method performance and may be difficult to characterize theoretically \cite{zhang2012}.  Simulation studies where the true data generating model is known are critical for evaluating performance of models that are commonly applied but that may not match the true generating model. 

We have created a comprehensive algorithm that makes use of an external data resource to generate an entire study where a right-censored outcome is a function of multiple correlated time-varying covariates of mixed type.  We have demonstrated that empirical estimates of the generated data are comparable to those of the external data source, suggesting the algorithm successfully captures the complexities of the data set.  In addition, we have provided open-source software that call existing R packages such as \emph{binnor}.  This approach allows other investigators to implement our algorithm, and further enhance it or tailor it for their own purposes.  Our algorithm captures multiple complexities of an existing longitudinal data set that includes correlation across features as well as correlation of a given feature within a subject or cluster.

There are limitations to our approach. Because variables containing proportions (bounded between 0 and 1) are generated from the normal distribution, further adjustments are necessary. Specifically, after generating data that describes whether a subject is ever exposed to a drug using marginal information, the algorithm includes steps to ensure individual proportions are between 0 and 1 as would be desired for such a variable. This intermediate step could have been avoided had we simulated from the Beta distribution. However, there is no existing way to simulate Beta random variables jointly with Binomially and Normally distributed random variables. In addition, generating categorical variables that contain more than 2 levels was challenging. We solved this by generating the categorical variable using a multinomial model after generating all binary and continuous characteristics.  An alternative approach is to stratify by levels of the covariate and simulate within each level. Our method is more desirable, however, in that it can be extended to accommodate additional categorical variables by fitting a series of multinomial models in a sequential fashion. When simulating a study, it may be desirable to use the existing covariates of a data set and only simulate the outcome, as the correlation structure of the features will be captured. However, that is not always feasible for two reasons: 1) The raw data may not be available to be fully used in the simulations; and 2) One may also need to vary parameters involved in the covariate generation.  Making use of an external data source may still be possible, however, by incorporating information about the marginal and joint distribution of key features in order to capture the essence of the complexities of the data set.  For example, if interested in methods for drawing causal inference in the presence of confounding, one could impose interrelationships among variables that represent confounding while maintaining marginal distributions of the characteristics.

Code for implementing our simulation tool, \textbf{SimulateCER}, is currently available. Effort to submit the R package to the Comprehensive R Archive Network (CRAN) is ongoing.  However, it can immediately be downloaded from github (https://github.com/qsuProjects/SimulateCER). We have created vignettes that demonstrate the use of the algorithm and these can be found here: https://github.com/qsuProjects/SimulateCER/tree/master/R/GENCOV/vignettes. In order to use the algorithm, there are a set of parameters that need to be defined and data sets that need to be provided with information related to each variable to be generated: (1) A matrix listing the type (ID variable, normal variable corresponding to the mean proportion for a binary time-varying variable, other normal variable, time-function continuous variable, binary time-varying variable, or binary static variable) of each variable and relevant parameters (e.g., the across-subject mean and variance and the within-subject variance for normal variables); (2) An across-subject correlation matrix; and, (3) A within-subject correlation matrix.

Our code recognizes if some components are not well-defined and adapts them so that data are generated as intended. For example, there are theoretical bounds on the maximum absolute correlation between a normal and a binomial variable depending on the population proportion for the latter. If the user specifies an implausible correlation for two such variables, the code will automatically adjust the correlation and provide a warning. Additionally, if the user does not provide within-subject variances for the normal variables, these will default to 1/3 of the corresponding across-subject variances. Finally, the code flexibly accommodates generation of data in arbitrary combinations of variable types. For example, any number of each variable type can be generated simultaneously (as long as there is at least one continuous and one binary variable). To omit one or more types of variables, the user can simply omit parameters for such variables in the parameters spreadsheet (e.g., time-function variables will not be generated in the absence of slope parameters).

We have demonstrated a simulation to determine power using our algorithm. Mainstream software packages that calculate power to address a variety of research questions in the presence of a large number of correlated features and small sample sizes do not exist. Currently, power calculation formulas allow the inclusion of a single correlation coefficient relating one exposure to other covariates. This only applies, however, to the case where there are time-invariant covariates and not multiple observations per subject. Simulation studies can be helpful here, where we can evaluate the power in the presence of multiple time-varying characteristics with varying correlation. In addition, we can evaluate the precision to assess predictive ability of a decision rule, the probability of identifying all relevant features in a molecular signature, the power of identifying at least one characteristic associated with outcome, and the power of identifying an exact number of characteristics that are associated with the outcome (with varying magnitudes of association). To facilitate access to the simulation-based power calculations, our team is developing a user-friendly interface using the Shiny package in R but code can already be accessed through github (https://github.com/qsuProjects/CERPower). Our team's original motivation for simulation scenarios, however, was to evaluate missing data approaches in the longitudinal setting in the presence of time-varying covariates and right-censored outcomes, which provides another example of use of such studies. 

In the example we looked at, from our power calculations that relied on our algorithm to simulate data, we found that we had 80\% power to detect HRs of 1.5 and higher if the exposure prevalence was 10\% or higher. The power was lower if the exposure prevalence was 5\%. Additionally, given 5 exposures that were associated with outcome (with varying magnitudes), we would very likely not be able to detect all of them, and we had only a 34\% probability of identifying 3 of the 5 drug exposures as being of interest. In contrast, the power calculations derived from \emph{stpower cox} could only describe the power of detecting one exposure given a certain association.  When prevalence of the exposure was 40\% and with a high level of clustering, \emph{stpower cox}'s estimated power was lower than what we obtained using our simulations but similar to the power obtained using \emph{stpower exponential}.  To summarize, there are three main differences between the simulation that demonstrated power and \emph{stpower cox}: 1) Using the command \emph{stpower} one cannot summarize statistics such as probability of detecting at least one or probability of detecting all 5; 2) The \emph{stpower cox} analysis assumes that each patient has one single record with no time-varying variables; and 3) While in \emph{stpower cox} one needs to summarize the correlation between exposure and covariates into one single value, that is not the case when using a simulation. 

We have developed an algorithm that allows simulation of data from a study that can capture multiple complexities of a real longitudinal data set.  Our algorithm performs well in that it can be used to closely resemble a desired external target and it can reflect complexities observed in real CER studies.  Further, it can easily be implemented using the open-source software we have provided in R.  We have shown that there are multiple uses of our algorithm to develop or evaluate methods involving right-censored outcomes that are functions of time-static and time-varying covariates that are correlated with each other and within a subject over time.    

\section*{Funding}
This work was supported by grants from the Patient-Centered Outcomes Research In-
stitute Grant ME-1303-5989.

\section*{Disclaimer}
The views expressed in this article are those of the authors and do not necessarily reflect the position or policy of the Patient-Centered Outcomes Research Institute or the United States government.

\section*{Notes}
All source code of the simulation study is available at the project's website: \\
\href{http://med.stanford.edu/qsu/projects/PCORI\_Home/code-in-progress.html}{\texttt{med.stanford.edu/qsu/projects/PCORI\_Home/code-in-progress.html}}

\section{Supplementary Material}

\includepdf[pages=-,pagecommand={},width=\textwidth]{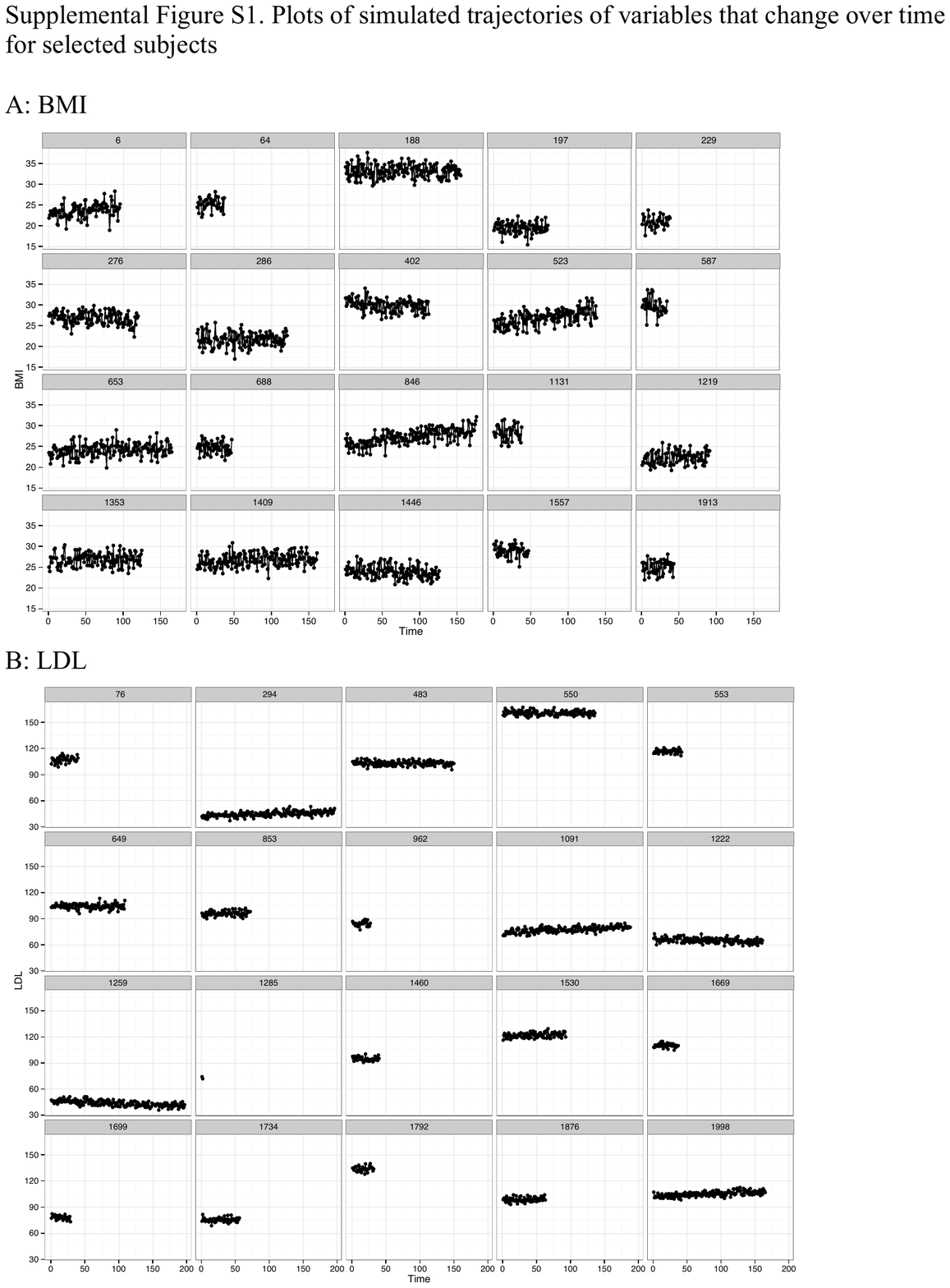}
\includepdf[landscape=true,pages=-,pagecommand={}]{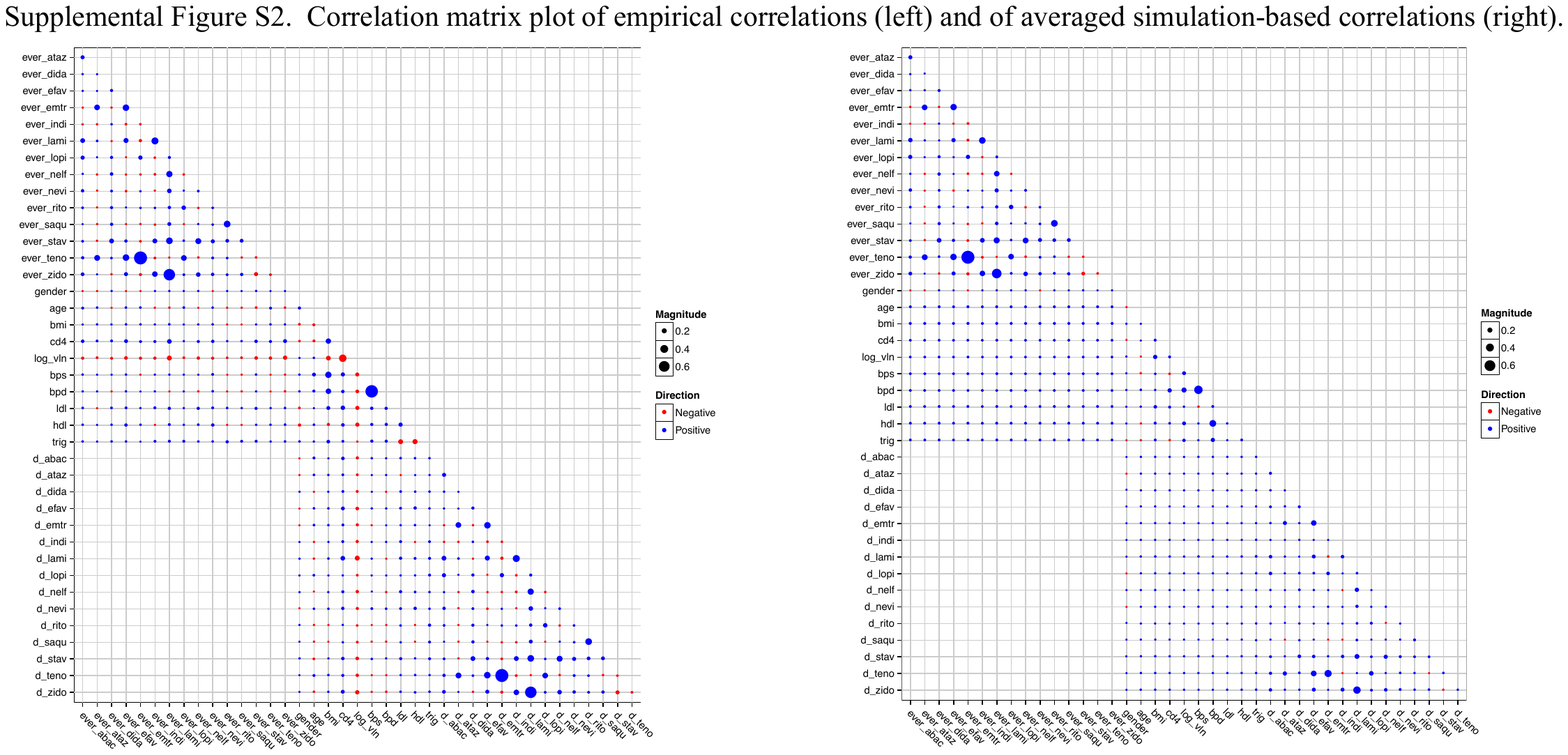}
\end{document}